\documentclass[sigconf]{acmart}

\AtBeginDocument{%
  \providecommand\BibTeX{{%
    \normalfont B\kern-0.5em{\scshape i\kern-0.25em b}\kern-0.8em\TeX}}}

\setcopyright{none}
\copyrightyear{}
\acmYear{}
\acmDOI{}

\acmConference[FAccTRec Workshop ’22]{4th FAccTRec Workshop on Responsible Recommendation}{18th-23rd September 2022}{Seattle, WA}
\acmPrice{}
\acmISBN{}

\setcopyright{none}

\begin{document}

\title{Towards Responsible Medical Diagnostics Recommendation Systems}

\author{Daniel Schlör}
\email{schloer@informatik.uni-wuerzburg.de}
\affiliation{%
  \institution{University of Würzburg}
  \streetaddress{Am Hubland}
  \city{Würzburg}
  \country{Germany}
}
\author{Andreas Hotho}
\email{hotho@informatik.uni-wuerzburg.de}
\affiliation{%
  \institution{University of Würzburg}
  \streetaddress{Am Hubland}
  \city{Würzburg}
  \country{Germany}
}





\maketitle


\section{Introduction}
The early development and deployment of hospital and healthcare information systems have encouraged the ongoing digitization of processes in hospitals.
Many of these processes, which previously required paperwork and telephone arrangements, are now integrated into IT solutions and require physicians and medical staff to interact with appropriate interfaces and tools.
Although this shift to digital data management and process support has benefited patient care in many ways, it requires physicians to accurately capture all relevant information digitally for billing and documentation purposes, which takes a lot of time away from actual patient care work. 
However, systematic collection of healthcare data over a long period of time offers opportunities to improve this process and support medical staff by introducing recommender systems.


In this position paper, we outline criteria for a responsible recommender system in the medical context from an application driven perspective and discuss potential design choices with a specific focus on accountability, safety, and fairness.




\section{Recommender Use Case}
In hospitals, several departments are specialized on specific diagnostics and offer services to other departments, such as radiology or laboratory work. To provide a diagnosis for a patient, the leading department of the case submits a request to another department requesting a specific examination. This examination is then performed at the department offering the diagnostic service and the results are reported back to the requesting department. 
For an efficient request process, the hospital decided to make the physician only define a rough diagnostic category such as wrist X-ray, rather than presenting the entire spectrum of about 2000 possible examinations, which can be performed and billed. However, for documentation and billing, the precise diagnostic procedure must be recorded. Therefore, the hospital has employed medical assistants who refine the request before passing it to the department. 

Incorporating recommender systems in this process can help to make this process more efficient. 
Therefore, they can be deployed in two process steps. First, as a support tool for medical assistants making the process of refining requests more efficient, and second, as a support tool for the requesting physicians, allowing them to request diagnostics more fine-grained and precise without overwhelming them with possible options and a time-consuming administrative process.

In addition to the process step, other technical considerations are relevant for the design of the recommender and subsequent potential ethical issues. Consider, for example, the question of what data base to train the recommender on. Diagnostics performed and billed have been recorded for a long time, as accounting requires them to be documented digitally and represent a relatively clean data basis, as, from requesting to performing examination, several experts were involved who correct potential issues. On the downside, many of these records cannot be connected to the original requests and their underlying decisions, as they might have been placed by phone in the past. 
On the other hand, the data available on the requesting side is mostly unstructured and limited to rather broad diagnostic categories. Additionally, the quality of the data is not ensured, since, for example, relevant proportions of requests are generated by inexperienced physicians which might be refined in subsequent process steps by experts for the respective diagnostics. 

Besides these technical considerations, developing and deploying a recommendation system in this sensitive medical context raises several ethical questions which will be addressed in the following sections.


\section{Responsibility}
The aspects of responsibility for the recommender system 
can be addressed from two views with contradicting outcomes. 
From a practical perspective, a recommender system that provides the foundation to request the required examination precisely reduces the workload for the assistant stuff with the potential to finally omit them in the process chain. From a socio-economic perspective, this relates to aspects of responsibility, as this could provide opportunities for the hospital to reduce staff in order to cut costs. 

From another perspective the same situation can have a different outcome. When the recommender allows a large number of standard cases to be requested directly, the saving of time is beneficial to the patients. The assistant staff can then focus on difficult cases, settle requests, which require further clarification, or be assigned to other tasks that overall improve care from the patient's perspective.

These perspectives suggest that the actual outcome for responsibility aspects depends on the decisions made by the hospital management, which is in line with the recent discussion presented by Gansky and Mcdonald \cite{gansky2022counterfacctual}, who remark that the organizational context in which the system is to be used need to be accounted. 



\section{Fairness}
Fairness of recommender systems can be evaluated from a procedural perspective \cite{lee2019procedural} with a focus on fairness within the decision process, or from an outcome driven perspective treating similar individuals or groups \cite{biega2018equity}.  

In their study, Tsuchiya and Dolan \cite{tsuchiya2009equality} show that 
the public majority prefers outcome centered fairness in a medical context. 
For the outcome centered perspective, however, the definition of the outcome is highly relevant. Obviously, the recommendation of an examination cannot be seen as outcome, since different (potentially protected) subgroups, for example, children and elderly require different examinations from a medical perspective. 
Additionally, learning from examinations performed in the past as ground truth does not necessarily reflect the best choice for patient and can not evaluate the success of subsequent treatments. A solution as proposed by Mei et al. \cite{mei2015outcome} using a cohort study-based evaluation
can be adopted to evaluate fairness. 
However, for our setting, the outcome remains difficult to define as even simple criteria, such as the waiting time until the diagnostics has been performed can be biased by group specific medical characteristics, such as urgency.

This shows that certain trade-offs with regard to fairness and ethics are necessary 
to reflect medical and economical reality 
\cite{beauchamp1994principles} and thereby to ensure acceptability for all stakeholders \cite{milano2020recommender}. Nevertheless, the audit of these trade-offs for their clinical and ethical validity have to be incorporated in the evaluation process of the recommender system.


\section{Accountability}

Accountability in the context of the examination recommender system boils down to who is, and who feels responsible for potential harm to patients caused by the complex socio-technical system including requesting physician, recommender system and their developers, medical assistants and executing physician \cite{eprs2022artificial}. A recent study on AI in healthcare of the European Parliamentary Research Service \cite{eprs2022artificial} identifies the need for ``new mechanisms and frameworks to ensure adequate accountability in medical AI''. 
While most proposed measures take a regulatory view, one measure suggests implementing processes to identify the roles of AI and users when AI-based decisions harm patients, making responsibilities explicit. 

As the physicians have the final decision to accept or reject recommendations, we propose to raise awareness for accountability and safety concerns as part of the requesting process by observing the acceptance rate of proposed recommendation. If this evaluation suggests that the requesting person relies on the recommendation to much, an awareness training step can be injected, e.g., by recommending invalid diagnostics which, if the user blindly accepts them, are stopped presenting a warning to the user.


\section{Transparency}
Transparency is identified as a key aspect for accountability and thus acceptability by Smith \cite{smith2021clinical}, who argues that for the use in clinical practice physicians have to account for their decision and will reject non-transparent AI systems as they cannot account for its outcome. On the other hand Clement et al. \cite{clement2021increasing} claim, that according to their empirical study, transparency favors the acceptance of low-quality recommendations, i.e., it introduces a behavioral change caused by backing model recommendations with explanations.

As the requests formulated by the physicians should be based on medical considerations, the recommender system in our setting primarily serves the purpose of improving the selection process rather than the decision process. 
A recommendation different from the actual intention must undergo a medical appraisal before qualifying as viable option. A automated rationalization or explanation of the recommendation might therefore shorten the thought process and thus undermine the necessary care. 
A less intrusive form of transparency, which provides the physician with the necessary information but without anticipating the actual reasoning can therefore be better suited. Providing the physician with similar cases to compare for a collaborative filtering based approach, or backing knowledge based recommendation with clinical guideline documents appear a promising intermediate path between opacity of the recommendations and fine-grained justifications with the potential to introduce unwanted behavioral changes.


\section{Compliance}
With regard to compliance, in Germany and the European Union, applicable regulations for medical AI include the \textit{2017/746 In Vitro Diagnostic Medical Devices Regulation (IVDR)} and the \textit{2017/745 Medical Devices Regulations (MDR)} \cite{eprs2022artificial}, while the latter is more applicable to the setting of diagnostics recommendation. Kiseleva \cite{kiseleva2020ai} concludes that this regulation can serve as initial legal framework, however it needs to be extended in terms of transparency and accountability which is extended by the need for risk assessment in the proposal of the European Parliamentary Research Service on Artificial intelligence in healthcare  \cite{eprs2022artificial}. 
From a practical perspective, the initiative of FUTURE-AI \cite{lekadir2021future} provides guidelines and best practices for trustworthy AI in medicine which should be taken into account for designing the recommender system.


\section{Safety}
The safety aspect is not only connected to the performance of the recommender itself, which is the obvious application evaluating how suitable the recommendations are. 
Instead safety can be also addressed from a user centric perspective as discussed for accountability, where misuse of a system can introduce safety issues, and from a data driven perspective, e.g., since the data most likely contains wrong data or noise. 
This is also reflected in practical guidelines and regulations as safety is best approached holistically, from data collection, annotation, over system design and evaluation, to the socio-technical and organizational context it is used \cite{lekadir2021future,dobbe2022system} and audited in this complex accordingly \cite{falco2021governing}.

For evaluating safety in our recommender system this means, that we can not only rely on performance metrics, especially as the data itself may be prone to errors and noise. While the performance of the system has to be in a range where the potential benefit outweighs the safety risks to be usable, the benefit and safety from an outcome perspective has to be consequently monitored especially during operation.


\section{Conclusion}

In this position paper, we introduced a recommender system for medical diagnostics recommendation in a user assisting context. We discussed implications of design decisions, the use case and the system with respect to responsibility, fairness, accountability, transparency, compliance and safety and toke a stand on possibilities the implications and issues could be addresses from a practical point of view. With this position paper outlining our recommender system, we hope to collect valuable input and participate in fruitful discussions at the FAccTRec Workshop towards designing and implementing recommendation in a more responsible way.

\section*{acknowledgement}
This research was supported by the Bavarian State Ministry for Science and
the Arts within the ``Digitalisierungszentrum für Präzisions- und Telemedizin''
(DZ.PTM) project, as part of the master plan ``BAYERN DIGITAL II''.





\bibliographystyle{ACM-Reference-Format}
\bibliography{sample-authordraft}




\end{document}